\documentclass[preprint,journal]{vgtc}       

\ifpdf
  \pdfoutput=1\relax                   
  \usepackage{graphicx}                
  \DeclareGraphicsExtensions{.pdf,.png,.jpg,.jpeg} 
\else
  \usepackage{graphicx}                
  \DeclareGraphicsExtensions{.eps}     
\fi%
\usepackage{amsmath} 
\usepackage[table]{xcolor}

\graphicspath{{figures/}{pictures/}{images/}{./}} 

\usepackage{microtype}                 
\PassOptionsToPackage{warn}{textcomp}  
\usepackage{textcomp}                  
\usepackage{mathptmx}                  
\usepackage{times}                     
\usepackage{cite}                      
\usepackage{tabu}                      
\usepackage{booktabs}                  

\onlineid{0}

\vgtccategory{Research}
\vgtcpapertype{application/design study}

\title{A Study on a User-Controlled Radial Tour for Variable Importance in High-Dimensional Data}
\author{Nicholas Spyrison, Dianne Cook, Kim Marriott}
\authorfooter{
\item
 Monash University\\
Australia
\\nicholas.spyrison@monash.edu\\
ORCiD: 0000-0002-8417-0212.
\item
 Monash University\\
Australia
\\dicook@monash.edu\\
ORCiD: 0000-0002-3813-7155
\item
 Monash University\\
Australia
\\kim.marriott@monash.edu\\
ORCiD: 0000-0002-9813-0377
}

\shortauthortitle{Spyrison \MakeLowercase{\textit{et al.}}: Study on User-Controlled Radial Tour}

\abstract{Principal component analysis is a long-standing go-to method for exploring multivariate data. The principal components are linear combinations of the original variables, ordered by descending variance. The first few components typically provide a good visual summary of the data. \emph{Tours} also make linear projections of the original variables but offer many different views, like examining the data from different directions. The grand tour shows a smooth sequence of projections as an animation following interpolations between random target bases. The manual radial tour rotates the selected variable's contribution into and out of a projection. This allows the importance of the variable to structure in the projection to be assessed. This work describes a mixed-design user study evaluating the radial tour's efficacy compared with principal component analysis and the grand tour. A supervised classification task is assigned to participants who evaluate variable attribution of the separation between two classes. Their accuracy in assigning the variable importance is measured across various factors. Data were collected from 108 crowdsourced participants, who performed two trials with each visual for 648 trials in total. Mixed model regression finds strong evidence that the radial tour results in a large increase in accuracy over the alternatives. Participants also reported a preference for the radial tour in comparison to the other two methods.
} 


\CCScatlist{ 
 \CCScat{10003120}{Human-centered computing}%
{}{};
 \CCScat{10003120.10003145.10011770}{Human-centered computing}{Visualization design and evaluation methods}{}
}

\vgtcinsertpkg

\begin{document}

\firstsection{Introduction}

\maketitle

%

Despite decades of research, multivariate data continues to provide fascinating challenges for visualization. Data visualization is important because it is a key element of exploratory data analysis \cite{tukey_exploratory_1977} for assessing model assumptions and as a cross-check on numerical summarization \cite{anscombe_graphs_1973, matejka_same_2017, yanai_hypothesis_2020}. One challenge is measuring if a new technique yields a more informed perception of information than current practices.

Dimension reduction is commonly used with visualization to provide informative low-dimensional summaries of quantitative multivariate data. Principal component analysis (PCA) \cite{pearson_liii._1901} is one of the first methods ever developed, and it remains very popular. Visualization of PCA is typically in the form of static scatterplots of a few leading components. When the scatterplot is accompanied by a visual representation of the basis they are called a biplot \cite{gabriel_biplot_1971}. A basis is a \(p \times d\) matrix of the linear combination of the \(p\) variables mapped to a smaller \(d\)-dimensional space. That is, it is an orthogonal rotation matrix, the magnitude, and the angle that the variables contribute.

Dynamic visualizations called \emph{tours} \cite{asimov_grand_1985} animate through a sequence of linear projections (orthonormal bases). Instead of a static view, tours provide a smoothly changing view by interpolating between bases. There are various types of tours distinguished by how the paths are generated. Asimov originally animated between randomly selected bases in the \emph{grand} tour. The \emph{manual} tour \cite{cook_manual_1997} allows for user control over the basis changes. A selected variable (or component) can be rotated into or out of view or to a particular value. The \emph{radial tour} \cite{spyrison_spinifex_2020} is a variant of the manual tour that fixes the contribution angle and changes the magnitude along the radius. The permanence of the data points from basis to basis holds information between intermediate interpolated projections, and the user control of the basis could plausibly lead to more information being perceived than a static display. This is a hypothesis that a user study could assess.

Empirical studies have rarely assessed tours. An exception is \cite{nelson_xgobi_1999}, who compares scatterplots of grand tours on 2D monitors with 3D (stereoscopic, not head-mounted) over \(n=15\) participants. Participants perform cluster detection, dimensionality estimation, and radial sparseness tasks on six-dimensional data. They find that stereoscopic 3D leads to more accuracy in cluster identification, though the time to interact with the display was much higher in the 3D environment. In this work, we extend the evaluation of tours which compares the radial tour as benchmarked against the grand tour and discrete pairs of principal components.

The contribution of this paper is an empirical user study comparing the radial tour against PCA and the grand tour for assessing variable attribution on clustered data. This is the first empirical evaluation of the radial or manual tour. We discuss how this fits with other multivariate data visualization techniques and coordinated views of linear projections.

\begin{figure*}
{\centering \includegraphics[width=1\linewidth]{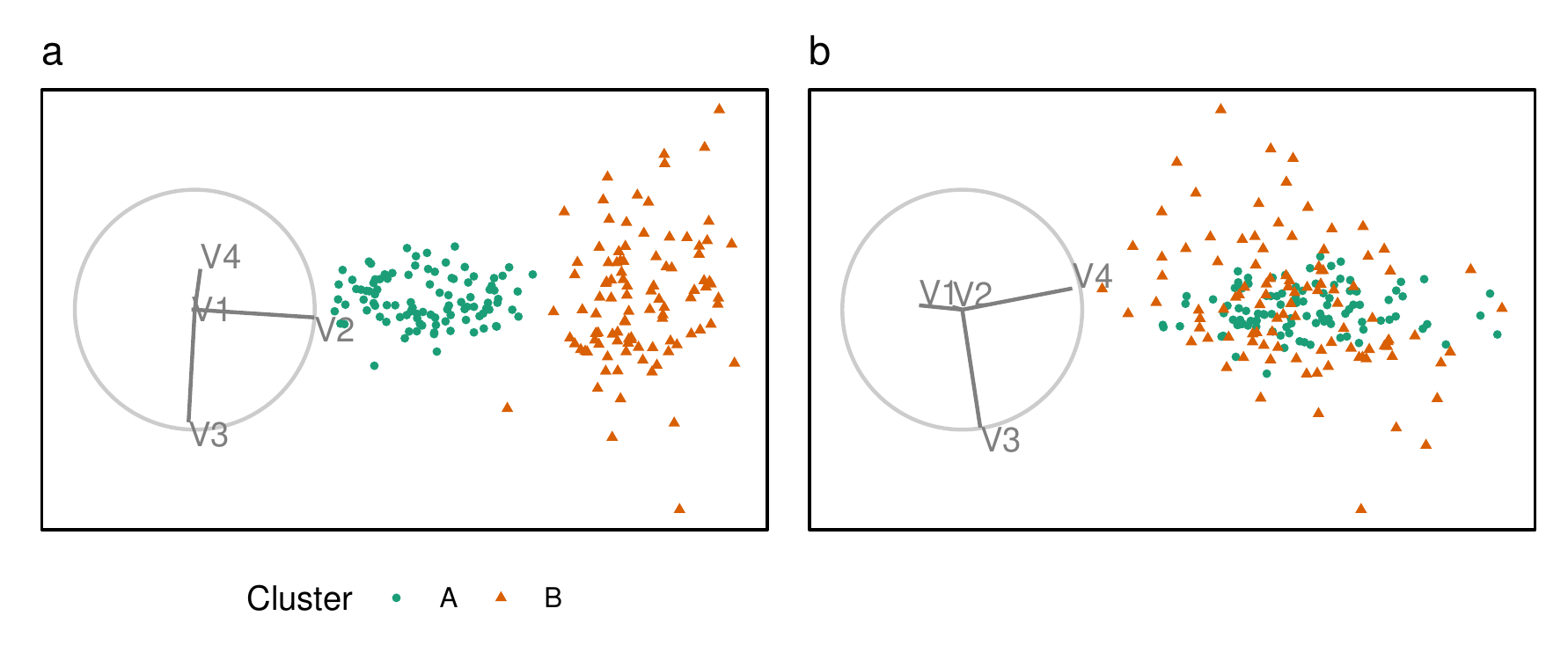}}
\caption{Illustration of cluster separation affected by variable importance. Panel (a) is a projection mostly of V2 and V3, and the separation between clusters is in the direction of V2, not V3. This suggests V2 is important for clustering, but V3 is not. Panel (b) shows a projection of mostly V3 and V4, with no contribution from V2 and little from V3. That there is no separation between the clusters indicates that V3 and V4 are not important.}\label{fig:figClSep}
\end{figure*}

We are particularly interested in assessing the effectiveness of the new radial tour relative to common practice with PCA and grand tour. The user influence over a basis, uniquely available in the radial tour, is crucial to testing variable sensitivity to the structure visible in projection. If the contribution of a variable is reduced and the feature disappears, then we say that the variable was sensitive to that structure. For example, \autoref{fig:figClSep} shows two projections of simulated data. Panel (a) has identified the separation between the two clusters. The contributions in panel (b) show no such cluster separation. The former has a large contribution of V2 in the direction of separation, while it is negligible in the right frame. Because of this, we say that V2 is sensitive to the separation of the clusters.

Variable sensitivity is important for the interpretation of machine learning models. They are the magnitude and direction of contribution to the model. It is important that developers maintain the interpretability of models. Exploratory Artificial Intelligence (XAI) \cite{adadi_peeking_2018, arrieta_explainable_2020}, is an emerging field that extends the interpretability of such black-box models. Multivariate data visualization is essential for exploring feature spaces and communicating interpretations of models \cite{biecek_dalex_2018, biecek_explanatory_2021, wickham_visualizing_2015}.

The paper is structured as follows. \autoref{sec:relatedwork} provides background on standard visualization methods and linear dimension reduction techniques. \autoref{sec:userstudy} describes the experimental factors, task, and accuracy measure used. The results of the study are discussed in \autoref{sec:results}. Conclusions and potential future directions are discussed in \autoref{sec:conclusion}. More results, participant demographics, and analysis of the response time are available in the Supplemental Materials.

\section{Related work}\label{sec:relatedwork}

Consider the data to be a matrix of $n$ observations (rows) and $p$ variables (columns), denoted as $X_{n \times p}$.

\subsection{Orthogonal multivariate visualization}

Grinstein \cite{grinstein_high-dimensional_2002} illustrates many multivariate visualization methods. In particular, this work shows examples of actual visuals. Liu \cite{liu_visualizing_2017} give a good classification and taxonomy of such methods. The content below focuses on the most common visuals that use the full data space before discussing linear combinations of those variables in projections.

\subsubsection{Scatterplot matrix}

\begin{figure}
{\centering \includegraphics[width=1\linewidth]{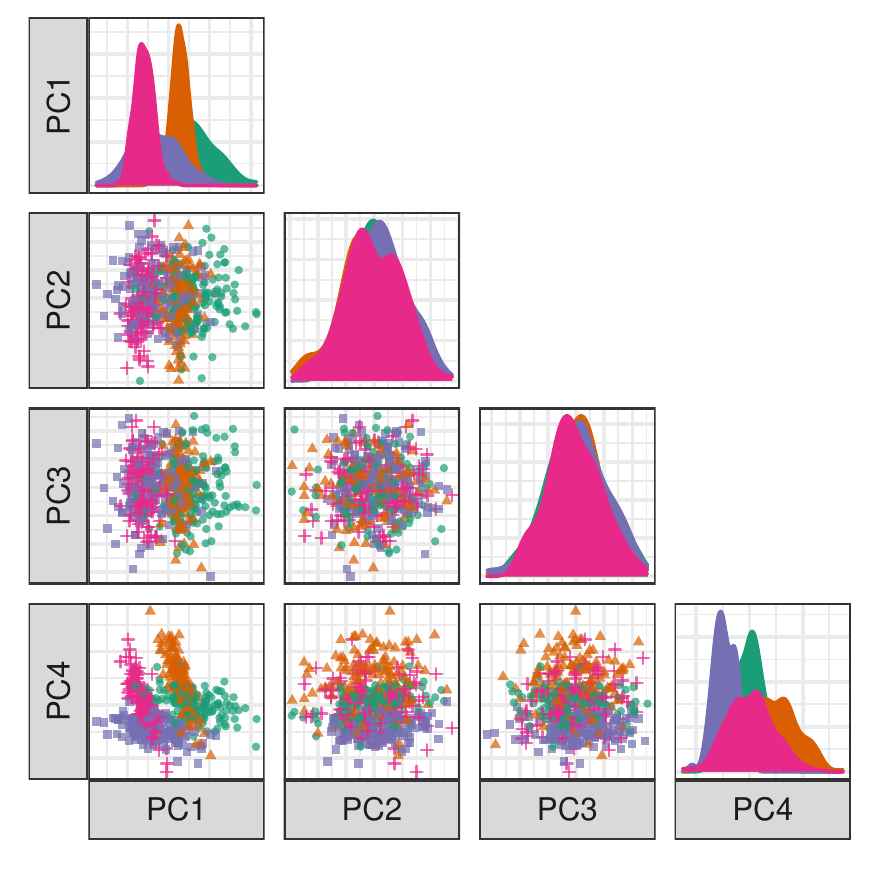}}
\caption{Scatterplot matrix of the first four principal components of 6D simulated data containing four classes. The separation between classes is primarily in PC1 and PC4. This is not uncommon because PCA is summarizing variance, not cluster structure.}\label{fig:figFactorPca}
\end{figure}

One could consider looking at $p$ histograms or univariate densities. Doing so will miss features in two or more dimensions. \autoref{fig:figFactorPca} shows a scatterplot matrix \cite{chambers_graphical_2018} of the four principal components of simulated data. Such displays do not scale well with dimensions because each plot would get less and less space. Scatterplot matrices can only display information in two orthogonal dimensions, so features in three dimensions may not be fully resolved.

\subsubsection{Parallel coordinates plot}

Another common way to display multivariate data is with a parallel coordinates plot \cite{ocagne_coordonnees_1885}. Parallel coordinates plots scale well with dimensions but poorly with observations as the lines overcrowd the display.
Parallel coordinate plots are asymmetric across variable ordering. In that, shuffling the order of the variable can lead to different conclusions. Another shortcoming is the graphical channel used to convey information. \cite{munzner_visualization_2014} suggests that position is the visual channel that is most perceptible to humans. In the case of parallel coordinates plots, the horizontal axes span variables rather than the values of one variable, causing the loss of a display dimension to be used by our most perceptible visual channel.

\subsection{Multivariate projections}

At some point, visualization will be forced to turn to dimension reduction to scale better with the dimensionality of the data. Below we introduce linear projections and the common principal component analysis. Then we touch on nonlinear projections and exclude them from consideration.

\subsubsection{Linear}

Let data, $X$, contain $n$ observations of $p$ variables. A linear projection maps a higher $p$-dimensional space onto a smaller $d$-space with an affine mapping (where parallel lines stay parallel). A projection, $Y$, is the resulting space of the data multiplied by a \emph{basis}, $A$, such that $Y_{n \times d} = X_{n \times p} \times A_{p \times d}$. This is essentially a reorientation of the original variable. This intuition is conveyed by thinking of a shadow as a 2D projection of a 3D object. Rotating the object changes the shadow it casts and, correspondingly, the basis that maps the reorientation of the object.

\subsubsection{Principal component analysis}

PCA is a good baseline of comparison for linear projections because of its frequent and broad use across disciplines. PCA \cite{pearson_liii._1901} defines new components, linear combinations of the original variables, ordered by decreasing variation through the help of eigenvalue matrix decomposition. While the resulting dimensionality is the same size, the benefit comes from the ordered nature of the components. The data can be said to be approximated by the first several components. The exact number is subjectively selected given the variance contained in each component, typically guided from a scree plot \cite{cattell_scree_1966}. Features with sizable signal regularly appear in the leading components that commonly approximate data. However, this is not always the case and component spaces should be fully explored to look for signal in components with less variation. This is especially true for cluster structure \cite{donnell1994}.

\subsubsection{Nonlinear}

Nonlinear transformations bend and distort spaces that are not entirely accurate or faithful to the original variable space. Popular modern methods include t-SNE and UMAP \cite{van_der_maaten_visualizing_2008, mcinnes_umap_2018}. There are various quality metrics, such as Trustworthiness, Continuity, Normalized stress, and Average local error, have been introduced to describe the distortion of the space \cite{espadoto_toward_2021, gracia_new_2016}. Unfortunately, these distortions are hard to visualize and comprehend, effectively breaking the variable interpretability of the resulting space. The intuition of this can be demonstrated with map projections. Snyder \cite{snyder_map_1987} lists over 200 different projections that distort the surface of the earth to display as a 2D map, each with unique properties and use cases.

Because of the difficulty of interpreting the distortions of nonlinear spaces and the added subjectivity of hyperparameter selection, we exclude nonlinear techniques and instead, decide to compare three linear techniques.

\subsection{Tours, animated linear projections} \label{sec:tours}

A \emph{tour} animates through many linear projections. One of the insightful features of the tour is the permanence of the data points; one can track the relative changes of observations as the basis changes, as opposed to discretely jumping to an orthogonal view angle with no intermediate information. Types of tours are distinguished by the generation of their basis paths \cite{lee_state_2021, cook_grand_2008}. In contrast with the discrete orientations of PCA, we compare continuous linear projection changes with grand and radial tours.

\subsubsection{Grand tours}

Target bases are selected randomly in a grand tour \cite{asimov_grand_1985}. These target bases are then geodesically interpolated for a smooth, continuous path. The grand tour is the first and most widely known tour. The random selection of target bases makes it a general unguided exploratory tool. The grand tour will make a good comparison that has a continuity of data points similar to the radial tour but lacks the user control enjoyed by PCA and radial tours.

\subsubsection{Manual and radial tours}

Whether an analyst uses PCA or the grand tour, cannot influence the basis. They cannot explore the structure identified or change the contribution of the variables. User-controlled steering is a key aspect of \emph{manual} tours that helps to test variable attribution.

\begin{figure*}
{\centering \includegraphics[width=1\linewidth]{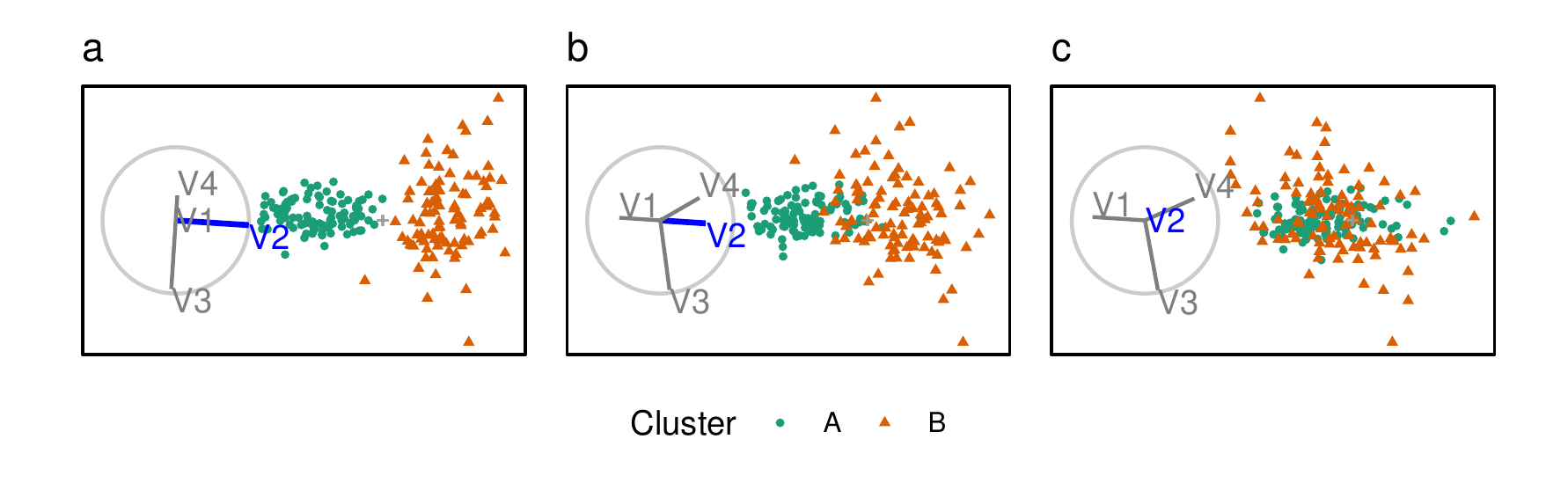}}
\caption{A radial tour changing the contribution of \texttt{V2}. The contribution is in the direction of cluster separation. When its contribution is removed, the clusters overlap (right). Because of this, we say that \texttt{V2} is sensitive to the separation of these two species.}\label{fig:figRadialTour}
\end{figure*}

The manual tour \cite{cook_manual_1997} defines its basis path by manipulating the basis contribution of a selected variable. A manipulation dimension is appended onto the projection plane, giving a full contribution to the selected variable. The target bases are then chosen to rotate this newly created manipulation space. This manipulation space is similarly orthogonally restrained. The data is projected through its interpolated basis and rendered into an animation. When the contribution of one variable changes, the contributions of the other variables must also change, to maintain the orthonormality of the basis. A key feature of the manual tour is that it allows users to control the variable contributions to the basis. Such manipulations can be queued in advance or selected in real time for human-in-the-loop analysis \cite{karwowski_international_2006}. Manual navigation is relatively time-consuming due to the vast volume of resulting view space and the abstract method of steering the projection basis. First, it is advisable to identify a basis of particular interest and then use the manual tour as a more directed, local exploration tool to explore the sensitivity of a variable's contribution to the feature of interest.

To simplify the task and keep its duration realistic, we consider a variant of the manual tour called a \emph{radial} tour. In a radial tour, the magnitude of along the radius with a fixed angle of contribution, as seen in \autoref{fig:figRadialTour}. The radial tour benefits from both continuity of the data alongside grand tours and user-steering via choosing the variable to rotate.

Manual tours have been recently made available in the \textbf{R} package \textbf{spinifex} \cite{spyrison_spinifex_2020}, which facilitates manual tour (and radial variant). It also provides an interface for a layered composition of tours and exporting to gif and mp4 with \textbf{gganimate} \cite{pedersen_gganimate_2020} or html widget with \textbf{plotly} \cite{sievert_interactive_2020}. It is also compatible with tours made by \textbf{tourr} \cite{wickham_tourr:_2011}. Now that we have a readily available means to produce various tours, we want to see how they fare against traditional discrete displays commonly used with PCA.

\subsection{Other animated linear projections}

The work of \cite{elmqvist_rolling_2008} allows users to interactively change the face of a local display by navigating to adjacent faces on a global overview scatterplot matrix. This offers analysts a way to geometrically explore the transition between adjacent faces of a scatterplot matrix as though rotating the face of dice at right angles. The interpolated bases between the orthogonal faces display linear combinations of three variables at varying degrees. This is what \cite{mcdonald1982} called a \emph{little tour} with the addition of user control. It is a particular type of manual tour where only horizontal or vertical rotation is allowed.

Star Coordinates \cite{kandogan_star_2000} also arrive at the biplot scatterplot displays starting from the perspective of radial parallel coordinates. \cite{lehmann_orthographic_2013} extend this idea, mapping it back to orthogonal projections. They provide a means to interpolate through PCA components, the orthogonal contributions of the scatterplot matrix, and the grand tour. This work also defines user-controlled interaction, similar to small steps in a manual or radial tour.

TripAdvisor \cite{nam_tripadvisornd_2012} is an interactive application that plans sequential interpolation between distant target bases. It also provides an additional global context of a subset of possible frames with glyph representation and an overview of variable attribution by summarizing the top ten principal components. It allows for user-steering by using a ``touchpad polygon''. This touchpad allows for contribution magnitudes to be changed. This is similar to an incremental change with the manual tour.

The number of orthogonal axes in static plots as well as the number of bases to view in a tour increase quadratically with the dimensions, $p$. This is why it is particularly important to properly select variables or otherwise reduce the dimensions before viewing. PCA, Linear discriminant analysis and entropy are common approaches to variable selection \cite{wang_linear_2017, sanchez_scaled_2018, sanchez_feature_2021}. Such methods often yield a sort of screeplot \cite{cattell_scree_1966} where the analyst selects a subjective, but informed, number of components to approximate the data while discarding the least information. The variable sensitivity we test for, in contrast, is the act of visual analysis of one variable's contribution to the structure. In practice, this is a tool for the analyst to fine-tune their variable selection or otherwise evaluate the resulting approximated space.

In order to further mitigate the view time, objective functions can be used to inform static or animated biplots. A dissimilarity statistic can be used to solve a basis path for showing a particularly interesting tour \cite{lehmann_optimal_2016}. More generally projection pursuit can be used to conduct a guided tour of any objective function applied to an embedding space \cite{cook_projection_1993, cook_grand_2008}. However, the function optimized is likely to show some feature of interest if it is ultimately selected by the analyst. The ability to stop and control the exploration at any point only stands to improve one's understanding of the data.

\subsection{Empirical evaluation}


Some studies compare visualizations across complete contributions of variables. Chang \cite{chang_evaluation_2018} conducted an $n=51$ participant study comparing parallel coordinate plots and scatterplot matrix either in isolation, sequentially, or as a coordinated view. Accuracy, completion time, and eye focus were measured for six tasks. Three tasks were more accurate with scatterplot matrix and three with parallel coordinates, while the coordinated view was usually marginally more accurate than the max of the separate visuals. Cao \cite{cao_z_glyph_2018} compare nonstandardized line-glyph and star-glyphs with standardized variants (with and without fill under the curve). Each of the $n=18$ participants performed 72 trials across the six visuals, two levels of dimensions, and two levels of observations. Visuals with variable standardization outperformed the nonstandardized variants, and the radial star-glyph reportedly outperformed the line variant.

Other studies have investigated the relative benefits of projecting to 2- or 3D scatterplots in PCA-reduced spaces. Gracia \cite{gracia_new_2016} conducted an $n=40$ user study comparing 2- and 3D scatterplots on traditional 2D monitors. Participants perform point classification, distance perception, and outlier identification tasks. The results are mixed and primarily have small differences. There is some evidence to suggest a lower error in distance perception from a 3D scatterplot. Wagner Filho \cite{wagner_filho_immersive_2018} performed an $n=30$ mixed-design study on PCA reduced space using scatterplot displays between 2D on monitors, 3D on monitors, and 3D display with a head-mounted display. None of the tasks on any dataset lead to a significant difference in accuracy. However, the immersive display reduced effort and navigation, resulting in higher perceived accuracy and engagement. Sedlmair \cite{sedlmair_empirical_2013} instead used two expert coders to evaluate 75 datasets and four dimension reduction techniques across the displays of 2D scatterplots, interactive 3D scatterplots, and 2D scatterplot matrices. They suggested a tiered guidance approach finding that 2D scatterplots are often sufficient to resolve a feature. If not, try 2D scatterplots on a different dimension reduction technique before going to scatterplot matrix display or concluding a true negative. They find that interactive 3D scatterplots help in very few cases.

\subsection{Conclusion}

Orthogonal axes visualizations either scale poorly with dimensionality or introduce an asymmetry of the variable ordering. Projections visualize the full $p$-data as fewer dimensions, traditionally 1-3 at a time. In linear, orthogonal projections, the resulting space is composed of a linear combination of the original variables that maintain variable interpretability. While nonlinear techniques distort and bend space in different ways that are hard to visualize and communicate.

Tours are linear projections that are animated over changes in the basis. Several more-recent, orthographic-star coordinate methods independently reach animated linear projections similar to tours. Some quality metrics and empirical studies compare techniques but scarcely with animated methods. Below we conduct a user study to compare the radial tour with PCA and the grand tour on a variable attribution task on clustered data.

\section{User study} \label{sec:userstudy}

The experiment was designed to assess the performance of the radial tour relative to the grand tour and PCA for interpreting the variable attribution to the separation between two clusters. Data were simulated across three experimental factors: location of the cluster separation, cluster shape, and data dimensionality. Participant responses were collected using a web application and crowdsourced through prolific.co, \cite{palan_prolific_2018} an alternative to MTurk.

\subsection{Objective} \label{sec:objective}

PCA will be used as a baseline for comparison as it is the most commonly used linear embedding. It will use static, discrete jumps between orthogonal components. The grand tour will act as a secondary control that will help evaluate the benefit of observation trackability between nearby animation bases but without user-control of its path. Lastly, the radial tour will be compared, which benefits from the continuity of animation and user control of the basis.

Then for some subset of tasks, we expect to find that the radial tour performs most accurately. Conversely, there is less to be certain about the accuracy of such limited grand tours as there is no objective function in selecting the bases; it is possible that the random selection of the target bases altogether avoids the bases showing cluster separation. However, given that the data dimensionality is modest, it is probable that the grand tour coincidentally regularly crossed bases with the correct information for the task.

Experimental factors and the definition of an accuracy measure are given below. The null hypothesis can be stated as:

\begin{align*}
  &H_0: \text{accuracy does not change across the visual methods} \\
  &H_\alpha: \text{accuracy does change across the visual methods}
\end{align*}

\subsection{Visual factors} \label{sec:standardization}

The visual methods are tested mixed design, with each visual being evaluated twice by each participant. Scatterplot matrices or parallel coordinates could alternatively be used to visualize these spaces. However, we opt to focus on single biplot displays to focus on the differences between the radial tour and its most comparable visuals, rather than a comprehensive comparison of visual methods. The rest of this section discusses the design standardization and unique input associated with each visual.

The visualization methods were standardized wherever possible. Data were displayed as 2D scatterplots with biplots. All aesthetic values (color-blind safe colors, shapes, sizes, absence of legend, and axis titles) were constant. The variable contribution biplot was always shown left of the scatterplot embeddings with their aesthetic values consistent. What did vary between visuals were their inputs.

\begin{figure}
{\centering \includegraphics[width=1\linewidth]{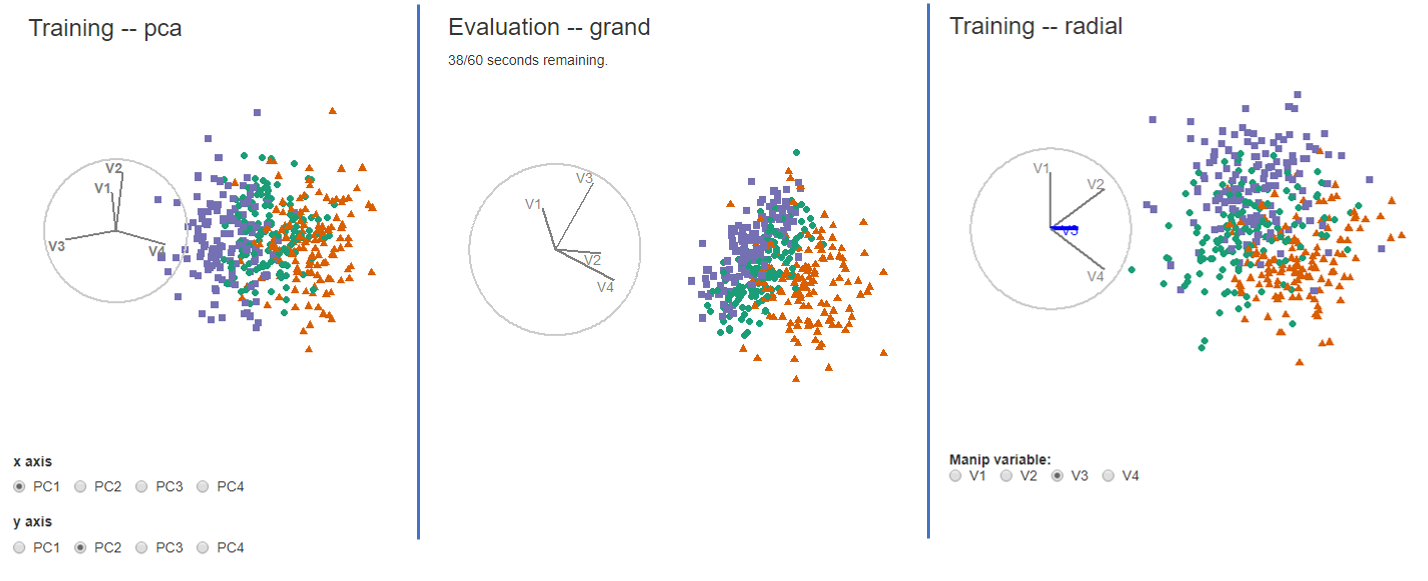}}
\caption{Examples of the application displays for PCA, grand tour, and radial tour.}\label{fig:figAppVis}
\end{figure}

PCA allowed users to select between the top four principal components for each axis regardless of the data dimensionality (four or six). Upon changing an axis, the visual would change to the new view of orthogonal components without displaying intermediate bases. There was no user input for the grand tour; users were instead shown a 15-second animation of the same randomly selected path (variables containing cluster separation were shuffled after simulation). Participants could view the same clip up to four times within the time limit. Radial tours allowed participants to select the manipulation variable. The starting basis was initialized to a half-clock design, where the variables were evenly distributed in half of the circle. This design was created to be variable agnostic while maximizing the independence of the variables. Selecting a new variable resets the animation where the new variable is manipulated to a complete contribution, zeroed contribution, and then back to its initial contribution. Animation and interpolation parameters were held constant across grand and radial tour (five bases per second with a step size of 0.1 radians between interpolated bases). \autoref{fig:figAppVis} displays screen captures of the visuals in the application.

\subsection{Experimental factors} \label{sec:expfactors}

\begin{figure}
{\centering \includegraphics[width=1\linewidth]{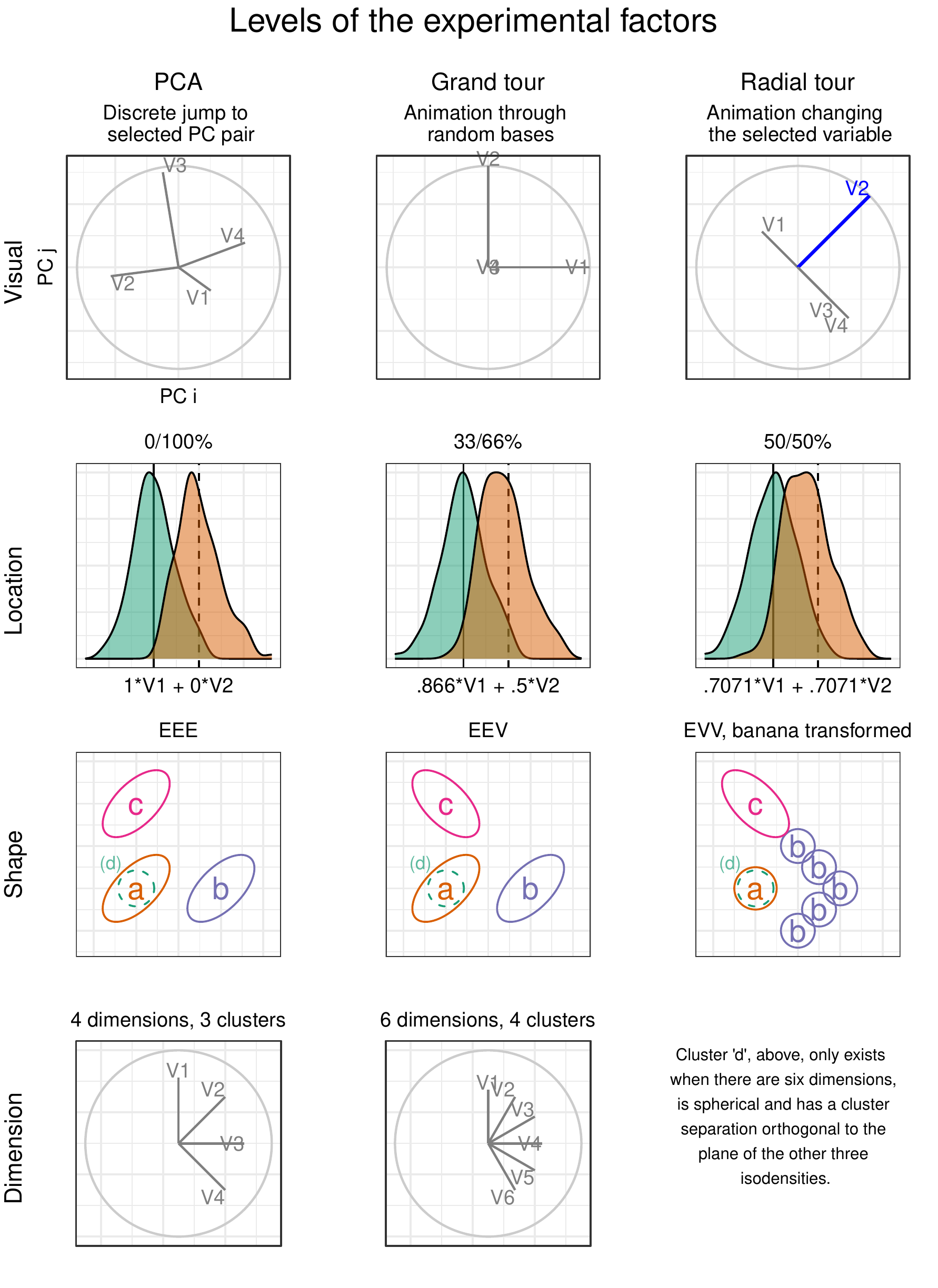}}
\caption{Levels of the visuals and three experimental factors: location of cluster separation, the shape of clusters, and dimensionality of the sampled data.}\label{fig:figExpFactors}
\end{figure}

In addition to the visual method, data are simulated across three experimental factors. First, the \emph{location} of the separation between clusters is controlled by mixing a signal and a noise variable at different ratios. Secondly, the \emph{shape} of the clusters reflects varying distributions of the data. And third, the \emph{dimension}-ality of the data is also tested. The levels within each factor are described below, and \autoref{fig:figExpFactors} gives a visual representation.

The \emph{location} of the separation between the clusters is at the heart of the measure. It would be good to test a few varying levels. To test the sensitivity, a noise and signal variable are mixed at different ratios. The separation between clusters is mixed at the following percentages: 0/100\% (not mixed), 33/66\%, 50/50\% (evenly mixed).

In selecting the \emph{shape} of the clusters, the convention given by Scrucca et al. (2016) is followed. They describe 14 variants of model families containing three clusters. The model family name is the abbreviation of the clusters' respective volume, shape, and orientation. The levels are either \emph{E}qual or \emph{V}ary. The models EEE, EEV, and EVV are used. For instance, in the EEV model, the volume and shape of clusters are constant, while the shape's orientation varies. The EVV model is modified by moving four-fifths of the data out in a ``\textgreater{}'' or banana-like shape.

\emph{Dimension}-ality is tested at two modest levels: four dimensions containing three clusters and six with four clusters. Such modest dimensionality is required to limit the difficulty and search space to make the task realistic for crowdsourcing.

\subsection{Task and evaluation} \label{sec:task}

With our hypothesis formulated, let us turn our attention to the task and how it is evaluated. Participants were asked to ``check any/all variables that contribute more than average to the cluster separation of the green circles and the orange triangles''. This was further explained in the explanatory video as ``mark any and all variable that carries more than their fair share of the weight, or one quarter in the case of four variables''. The participant instruction video can be viewed at \url{https://vimeo.com/712674984}.

The instructions iterated several times in the video were: 1) use the input controls to find a basis that contains separation between the clusters of green circles and orange triangles, 2) look at the orientation of the variable contributions in the grey circle (biplot axes orientation), and 3) select all variables that contribute more than uniformed distributed cluster separation in the scatterplot. Independent of the experimental level, participants were limited to 60 seconds for each evaluation of this task. This restriction did not impact many participants as the 25th, 50th, and 75th quantiles of the response time were about 7, 21, and 30 seconds, respectively.

The accuracy measure of this task was designed with a couple of features in mind. 1) Symmetric about the expected value, without preference for under- or over-guessing. 2) Heavier than linear weight with an increasing difference from the expected value. The following measure is defined for evaluating the task.

Let the data \(\textbf{X}_{ijk}, i=1,...,n; j=1,...,p; k=1,...,K\) be simulated observations containing clusters of observations of different distributions. Where \(n\) is the number of observations, \(p\) is the number of variables, and \(K\) indicates the number of clusters. Cluster membership is exclusive; an observation cannot belong to more than one cluster.

\begin{figure*}
{\centering \includegraphics[width=1\linewidth]{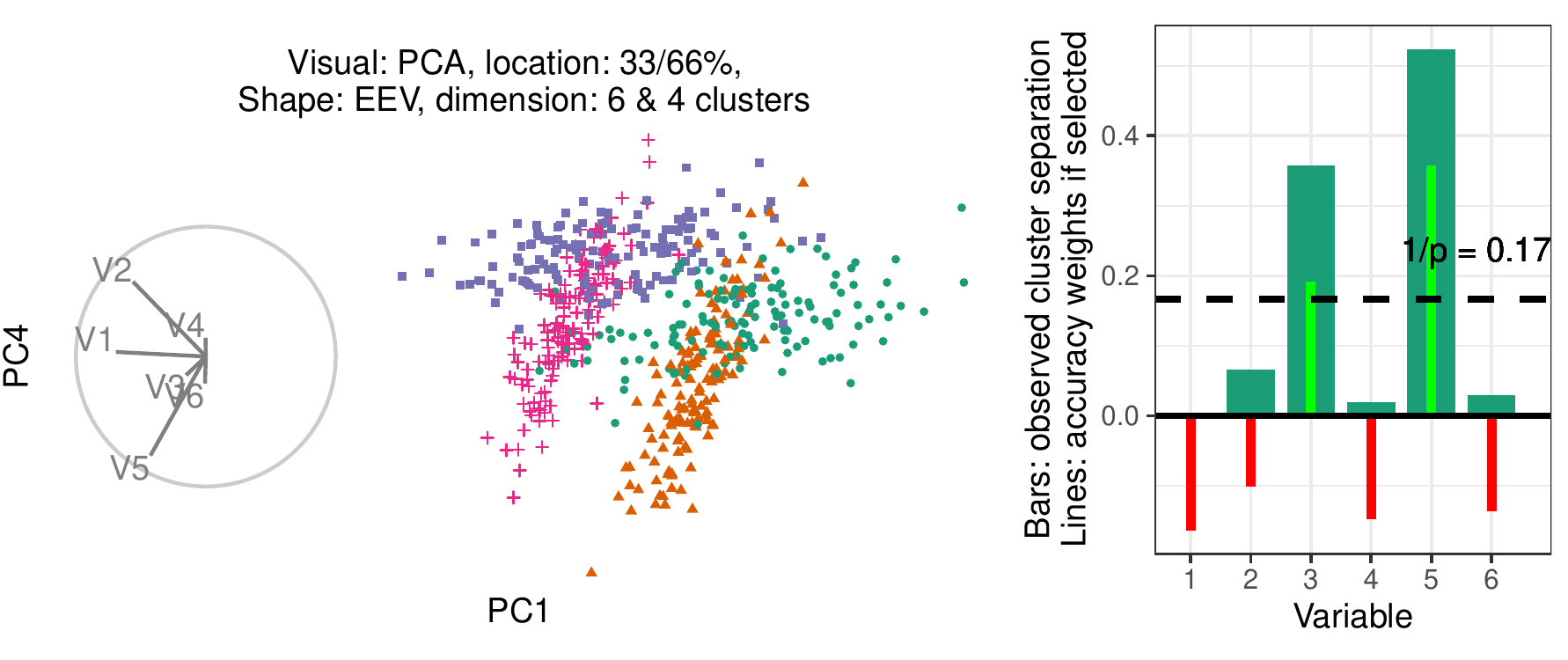}}
\caption{Illustration of how accuracy is measured. (L), Scatterplot and biplot of PC1 by PC4 of a simulated data set (R) illustrate cluster separation between the green circles and orange triangles. Bars indicate observed cluster separation, and (red/green) lines show the accuracy of the variable if selected. The horizontal dashed line has a height $1 / p$, the expected value of cluster separation. The accuracy weights equal the signed square of the difference between each variable value and the dashed line.}\label{fig:figBiplotScoring}
\end{figure*}

The weights, \(w\), is a vector of the variable-wise difference between the mean of two clusters of less \(1/p\), the expected cluster separation if it were uniformly distributed. Accuracy, \(A\) is defined as the signed square of these weights if selected by the participant. Participant responses are a logical value for each variable --- whether or not the participant thinks each variable separates the two clusters more than uniformly distributed separation. Weights comparing clusters 1 and 2 are calculated as follows:

\begin{align*}
  A =&~~ \sum_{j=1}^{p}I(j) \cdot sign(w_j) \cdot w_j^2, \mbox{~where~~~~~~~~~~~~~~} \\
  w_{j} =&~~ \frac{
  \overline{X}_{\cdot j1} - \overline{X}_{\cdot j2}
  }
  {\sum_{j=1}^{p}(|\overline{X}_{\cdot j1} - \overline{X}_{\cdot j2}|)} - \frac{1}{p}
\end{align*}

\[
\begin{array}{l}
\text{where} \\
~~I \text{ is the indicator function, returning a binary response}
\\
~~\overline{X}_{\cdot j k} \text{, mean of the } j\text{-th variable of the } k\text{-th cluster}
\end{array}
\]

\noindent where \(I(j)\) is the indicator function, the binary response for variable \(j\). \autoref{fig:figBiplotScoring} shows one projection of a simulation with its observed variable separation (wide bars), expected uniform separation (dashed line), and accuracy if selected (thin vertical lines).

\subsection{Randomized factor assignment}

Now, with simulation and their artifacts in hand, this section covers how the experimental factors are assigned and demonstrate how this is experienced from the participant's perspective.

The study is sectioned into three periods. Each period is linked to a randomized level of visual and location. The order of dimension and shape are of secondary interest and are held constant in increasing order of difficulty; four then six dimensions and EEE, EEV, then EVV-banana, respectively.

Each period starts with an untimed training task at the simplest remaining experimental levels; location = 0/100\%, shape = EEE, and four dimensions with three clusters. This serves to introduce and familiarize participants with input and visual differences. After the training, the participant performs two trials with the same visual and location level across the increasing difficulty of dimension and shape. The plot was removed after 60 seconds, though participants rarely reached this limit.

We assigned these factors based on the following order: visual methods, location, shape, and dimensionality. We first assigned three visual methods to three different sessions. The session order and the order of location follow a nested Latin square. The order of dimension and shape are assigned based on increasing order of difficulty.

Through pilot studies sampled by convenience (information technology and statistics Ph.D.~students attending Monash University), it was estimated that three complete evaluations are needed to power the study properly, a total of \(N = 3 \times 3!^2 = 108\) participants.

\subsection{Participants} \label{sec:articipants}

\(N = 108\) participants were recruited via prolific.co (Palan and Schitter 2018). Participants are restricted based on their claimed education requiring that they have completed at least an undergraduate degree (some 58,700 of the 150,400 users at the time). This restriction is used on the premise that linear projections and biplot displays will not be regularly used for consumption by general audiences. There is also the implicit filter that Prolific participants must be at least 18 years of age and have implicit biases of timezone, internet availability, language compatibility, and socioeconomic status. Participants were compensated for their time at \pounds 7.50 per hour, whereas the mean duration of the survey was about 16 minutes. Previous knowledge or familiarity was minimal, as validated in the follow-up survey. The Supplemental Materials include a heatmap distribution of age and education paneled across preferred pronouns of the participants that completed the survey, who are relatively young, well-educated, and slightly more likely to identify as males.

\section{Results} \label{sec:results}

To recap, the primary response variable is accuracy, as defined in \autoref{sec:task}. Two primary data sets were collected; the user study evaluations and the post-study survey. The former is the 108 participants with the experimental factors: visual, location of the cluster separation signal, the shape of the variance-covariance matrix, and the dimensionality of the data. Experimental factors and randomization were discussed in \autoref{sec:expfactors}. A follow-up survey was completed by 84 of these 108 people. It collected demographic information (preferred pronoun, age, and education) and subjective measures for each visual (preference, familiarity, ease of use, and confidence).

Below a battery of mixed regression models is built to examine the degree of the evidence and the size of the effects of the experimental factors. Then, Likert plots and rank-sum tests to compare the subjective measures between the visuals.

\subsection{Accuracy}

To quantify the contribution of the experimental factors to the accuracy, mixed-effects models were fit. All models have a random effect term on the participant and the simulation. These terms explain the amount of error attributed to the individual participant's effect and variation due to the random sampling data.

In building a set of models to test, a base model with only the visual term being compared with the full linear model term and progressively interacting with an additional experimental factor. The models with three and four interacting variables are rank deficient; there is not enough varying information in the data to explain all interacting terms.

\begin{small}
$$
\begin{array}{ll}
\textbf{Fixed effects} &\textbf{Full model} \\
\alpha                 &\widehat{Y} = \mu + \alpha_i + \textbf{Z} + \textbf{W} + \epsilon \\
\alpha + \beta + \gamma + \delta &\widehat{Y} = \mu + \alpha_i + \beta_j + \gamma_k + \delta_l + \textbf{Z} + \textbf{W} + \epsilon \\
\alpha \times \beta + \gamma + \delta &\widehat{Y} = \mu + \alpha_i \times \beta_j + \gamma_k + \delta_l + \textbf{Z} + \textbf{W} + \epsilon \\
\alpha \times \beta \times \gamma + \delta &\widehat{Y} = \mu + \alpha_i \times \beta_j \times \gamma_k + \delta_l + \textbf{Z} + \textbf{W} + \epsilon \\
\alpha \times \beta \times \gamma \times \delta &\widehat{Y} = \mu + \alpha_i \times \beta_j \times \gamma_k \times \delta_l + \textbf{Z} + \textbf{W} + \epsilon
\end{array}
$$

$$
\begin{array}{l}
\text{where} \\
~~\mu \text{ is the intercept of the model} \\
~~\alpha_i \text{ is the visual}~|~i\in (\text{pca, grand, radial}) \\
~~\beta_j  \text{ is the location}~|~j\in (\text{0/100, 33/66, 50/50\% mix}) \\
~~\gamma_k \text{ is the shape}~|~k\in (\text{EEE, EEV, EVV banana}) \\
~~\delta_l \text{ is the dimension}~|~l\in \text{(4 \& 3, 6 \& 4) var \& clusters} \\
~~\textbf{Z} \sim \mathcal{N}(0,~\tau) \text{ is the random effect of participant} \\
~~\textbf{W} \sim \mathcal{N}(0,~\upsilon) \text{ is the random effect of simulation} \\
~~\epsilon \sim \mathcal{N}(0,~\sigma) \text{ is the remaining error of the model}
\end{array}
$$
\end{small}

\begin{table}

\caption{\label{tab:marksCompTbl}Model performance of random effect models regressing accuracy. Complex models perform better in terms of $R^2$ and RMSE, yet AIC and BIC penalize their large number of fixed effects in favor of the much simpler model containing only the visuals. Conditional $R^2$ includes error explained by the random effects, while marginal does not.}
\centering
\fontsize{8}{10}\selectfont
\begin{tabular}[t]{lrrrrr}
\toprule
Model & AIC & BIC & R2 cond. & R2 marg. & RMSE\\
\midrule
\cellcolor{gray!6}{a} & \cellcolor{gray!6}{\textbf{-71}} & \cellcolor{gray!6}{\textbf{-71}} & \cellcolor{gray!6}{-44.219} & \cellcolor{gray!6}{0.303} & \cellcolor{gray!6}{0.289}\\
a+b+c+d & -45 & -45 & 4.063 & 0.334 & 0.294\\
\cellcolor{gray!6}{a*b+c+d} & \cellcolor{gray!6}{-26} & \cellcolor{gray!6}{-25} & \cellcolor{gray!6}{41.445} & \cellcolor{gray!6}{0.338} & \cellcolor{gray!6}{0.293}\\
a*b*c+d & 28 & 32 & 167.092 & \textbf{0.383} & 0.309\\
\cellcolor{gray!6}{a*b*c*d} & \cellcolor{gray!6}{105} & \cellcolor{gray!6}{116} & \cellcolor{gray!6}{\textbf{360.052}} & \cellcolor{gray!6}{0.37} & \cellcolor{gray!6}{\textbf{0.19}}\\
\bottomrule
\end{tabular}
\end{table}

\begin{table}

\caption{\label{tab:marksCoefTbl}The task accuracy model coefficients for $\widehat{Y} = \alpha \times \beta + \gamma + \delta$, with visual = pca, location = 0/100\%, shape = EEE, and dim = 4 held as baselines. Visual being radial is the fixed term with the strongest evidence supporting the hypothesis. Interacting with the location term, there is evidence suggesting radial performs with minimal improvement for 33/66\% location mixing.}
\centering
\fontsize{8}{10}\selectfont
\begin{tabular}[t]{lrrrrrl}
\toprule
  & Est & SE & df & t val & Prob & \\
\midrule
\cellcolor{gray!6}{(Intercept)} & \cellcolor{gray!6}{0.10} & \cellcolor{gray!6}{0.06} & \cellcolor{gray!6}{16.1} & \cellcolor{gray!6}{1.54} & \cellcolor{gray!6}{0.143} & \cellcolor{gray!6}{}\\
\addlinespace[0.3em]
\multicolumn{7}{l}{\textbf{Factor}}\\
\hspace{1em}VisGrand & 0.06 & 0.04 & 622.1 & 1.63 & 0.104 & \\
\cellcolor{gray!6}{\hspace{1em}VisRadial} & \cellcolor{gray!6}{0.14} & \cellcolor{gray!6}{0.04} & \cellcolor{gray!6}{617.0} & \cellcolor{gray!6}{3.77} & \cellcolor{gray!6}{0.000} & \cellcolor{gray!6}{***}\\
\addlinespace[0.3em]
\multicolumn{7}{l}{\textbf{Fixed effects}}\\
\hspace{1em}Loc33/66\% & -0.02 & 0.07 & 19.9 & -0.29 & 0.777 & \\
\cellcolor{gray!6}{\hspace{1em}Loc50/50\%} & \cellcolor{gray!6}{-0.04} & \cellcolor{gray!6}{0.07} & \cellcolor{gray!6}{20.0} & \cellcolor{gray!6}{-0.66} & \cellcolor{gray!6}{0.514} & \cellcolor{gray!6}{}\\
\hspace{1em}ShapeEEV & -0.05 & 0.06 & 11.8 & -0.82 & 0.427 & \\
\cellcolor{gray!6}{\hspace{1em}ShapeBanana} & \cellcolor{gray!6}{-0.09} & \cellcolor{gray!6}{0.06} & \cellcolor{gray!6}{11.8} & \cellcolor{gray!6}{-1.54} & \cellcolor{gray!6}{0.150} & \cellcolor{gray!6}{}\\
\hspace{1em}Dim6 & -0.01 & 0.05 & 11.8 & -0.23 & 0.824 & \\
\addlinespace[0.3em]
\multicolumn{7}{l}{\textbf{Interactions}}\\
\cellcolor{gray!6}{\hspace{1em}VisGrand:Loc33/66} & \cellcolor{gray!6}{-0.02} & \cellcolor{gray!6}{0.06} & \cellcolor{gray!6}{588.9} & \cellcolor{gray!6}{-0.29} & \cellcolor{gray!6}{0.774} & \cellcolor{gray!6}{}\\
\hspace{1em}VisRadial:Loc33/66 & -0.12 & 0.06 & 586.5 & -2.13 & 0.033 & *\\
\cellcolor{gray!6}{\hspace{1em}VisGrand:Loc50/50} & \cellcolor{gray!6}{-0.03} & \cellcolor{gray!6}{0.06} & \cellcolor{gray!6}{591.6} & \cellcolor{gray!6}{-0.47} & \cellcolor{gray!6}{0.641} & \cellcolor{gray!6}{}\\
\hspace{1em}VisRadial:Loc50/50 & -0.06 & 0.06 & 576.3 & -1.16 & 0.248 & \\
\bottomrule
\end{tabular}
\end{table}

\autoref{tab:marksCompTbl} compares the model summaries across increasing complexity. The \(\alpha \times \beta + \gamma + \delta\) model is selected to examine in more detail as it has relatively high condition \(R^2\) and not overly complex interacting terms. \autoref{tab:marksCoefTbl} looks at the coefficients for this model. There is strong evidence suggesting a relatively large increase in accuracy from the radial tour, though there is evidence that almost all of the increase the is lost under 33/66\% mixing.

\begin{figure}
{\centering \includegraphics[width=1\linewidth]{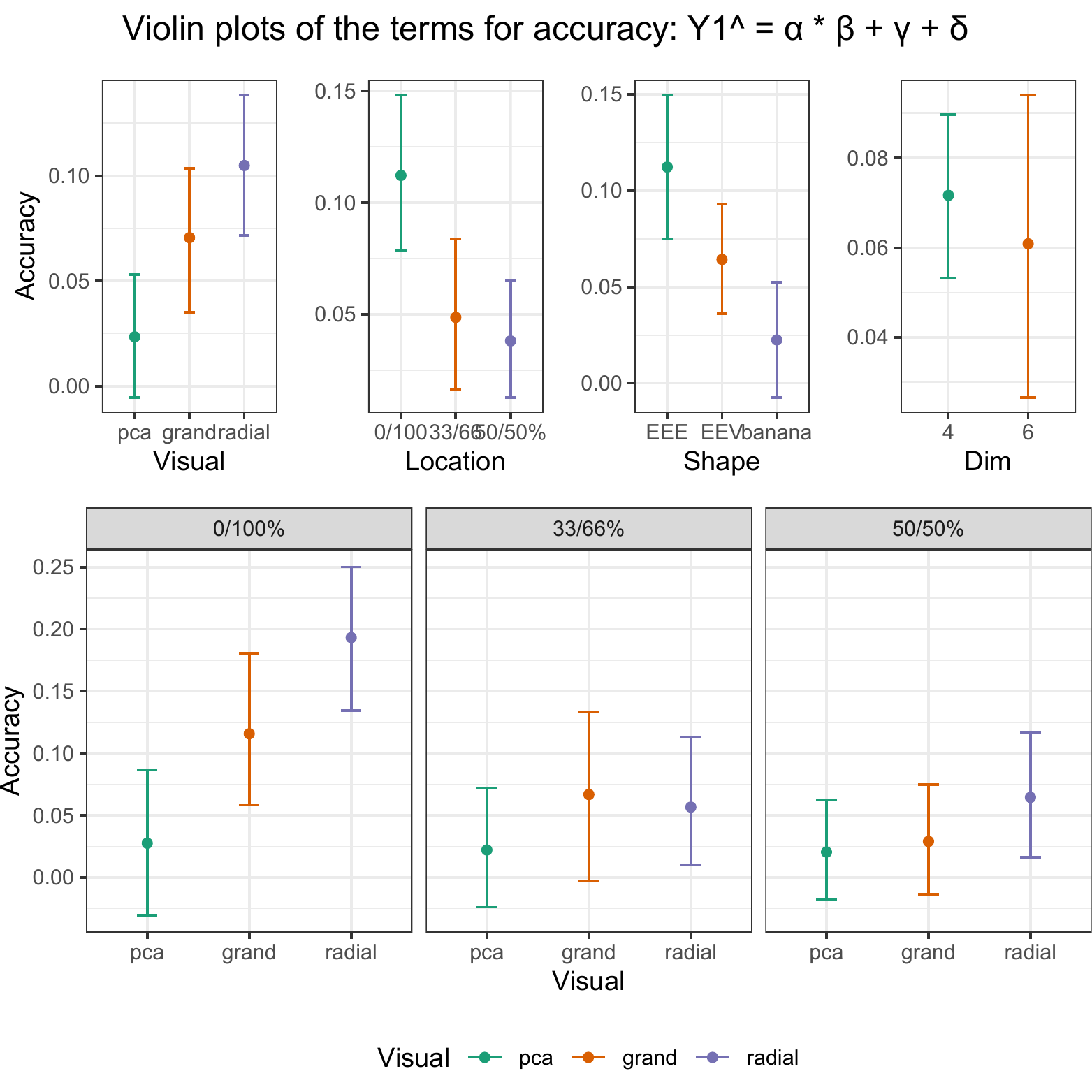} }
\caption{Accuracy of terms of the model $\widehat{Y} = \alpha \times \beta + \gamma + \delta$. Viewing the marginal accuracy of the terms corroborates the primary findings that the use of the radial tour leads to a significant increase in accuracy, at least over PCA, and this effect is particularly well supported when no location mixing is applied.}\label{fig:figMarksABcd}
\end{figure}

We also want to visually examine the conditional variables in the model. \autoref{fig:figMarksABcd} illustrates the  accuracy for each model term shown as mean and 95\% confidence interval.

\subsection{Subjective measures}\label{subjective-measures}

Modeling has proven that the use of the radial tour leads to a sizable improvement in the accuracy measure for this task. This is not the whole story. It is desirable to know what the users think of using the visuals. We follow the direction set by \cite{wagner_filho_immersive_2018}. They observe four subjective measures. The following were used in this study: confidence, ease of use, prior familiarity, and preference. Each of these questions was asked of all for each visual as 5-point Likert items.

\begin{figure}
{\centering \includegraphics[width=1\linewidth]{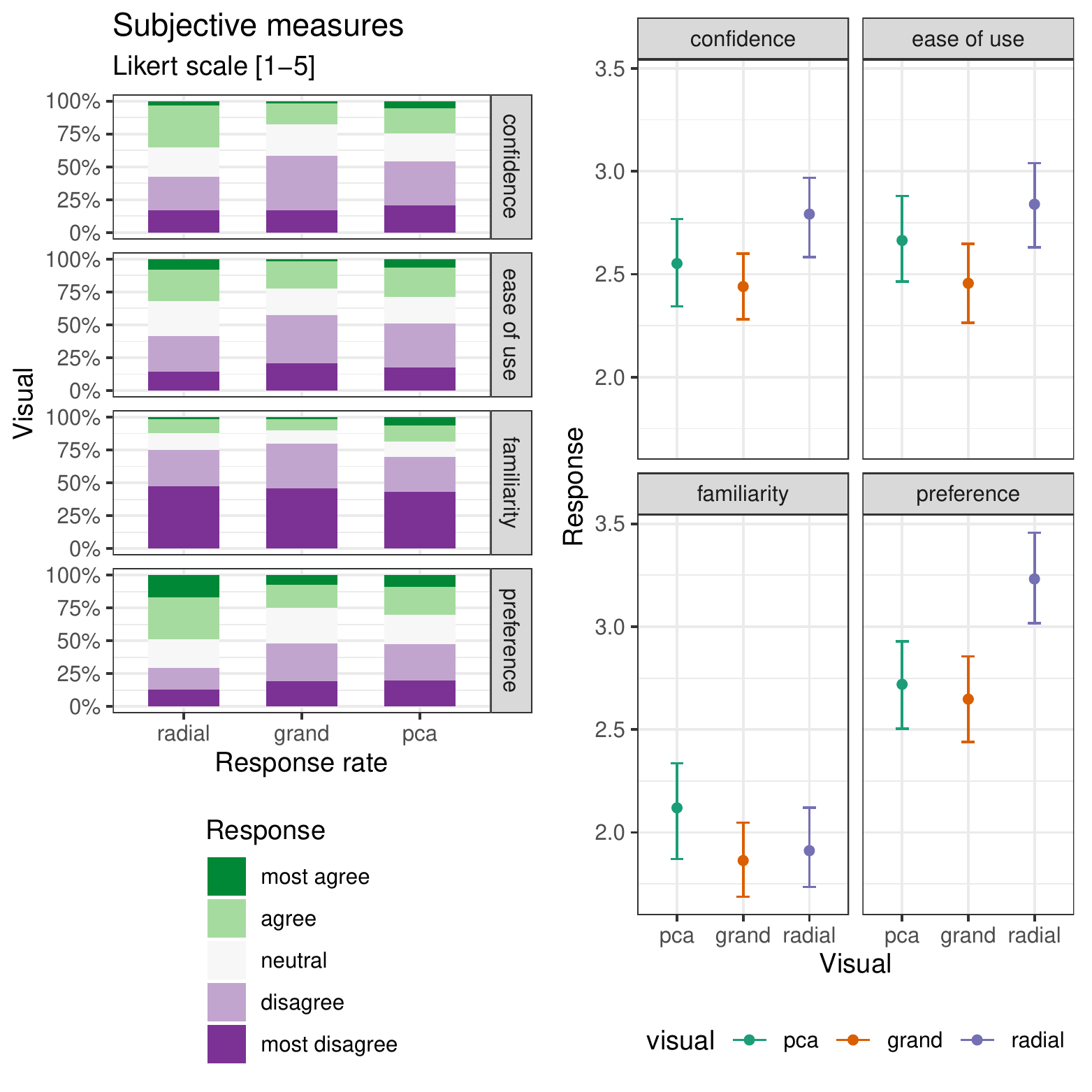}}
\caption{The subjective measures of the 84 responses of the post-study survey with five-point Likert items levels of agreement. (L) Likert plots (stacked percent bar plots) with (R) mean and 95\% CI of the same measures. Participants are more confident using the radial tour and find it easier to use than the grand tour. The radial tour is the most preferred visual.}\label{fig:figSubjectiveMeasures}
\end{figure}

The 84 evaluations of the post-study survey are shown in \autoref{fig:figSubjectiveMeasures}. The figure uses Likert plots or stacked percentage bar plots and asscoisated mean and 95\% confidence intervals.

There was strong evidence to support that participants preferred the radial tour to either alternative. There is less evidence that the radial tour led to more confidence and was found easier to use than the grand tour. In confirmation of expectations, crowdsourced participants had low familiarity with all visuals, with no difference in mean supported.

\section{Discussion}\label{discussion}

Data visualization is an integral part of understanding relationships in data and how models are fitted. When it comes to multivariate data giving a comprehensive view quickly becomes difficult as the dimensions become sizable. Analysts have the task of choosing which visualization technique to use. Because the viewing volume/time of multivariate spaces typically increase quadratically with dimensions dimension reduction must be properly conducted. While there are optimization methods for static and animated visuals, the particular function used is a guided choice of the analyst.

\autoref{sec:relatedwork} discussed various types of visualization which are may be preferred for differing tasks and ends. The visualization and perception of multivariate spaces is a broad and heterogeneous task. This work focuses a subset of linear projections and especially sheds light on potential benefit of providing user control in conjunction with the animated projection over many bases as a radial tour.

The radial tour is a method for the analyst to choose a variable to alter its contribution to the basis. The animation over small changes to the basis allows the sensitivity of the structure to be assessed from the variable contribution. The hypothesis is that user control over the basis and the permanence of observations between intermediate frames may lead to a better perception of the variable attribution causing the separation of clusters.

A mixed modeling analysis of the study provides strong support for this conclusion. That is, there is significant evidence to suggest the use of the radial tour leads to a sizable increase in accuracy. One unexpected caveat is that mixing the location of the signal at 33/66\% almost completely negates this gain. Perhaps this is because the ``half-clock'' basis used did not give enough weight to the variable containing the small fraction. It was also interesting to note that no level of the experimental factors alone had a significant effect on this setup. Lastly, the follow-up survey asked participants to evaluate measures of the visuals. Most notably, participants preferred the radial tour to the other visuals. Knowing that the radial tour outperforms alternatives and is the preferred choice can help inform the selection of visual methods for developers and analysts.

There are several implicit limitations to this study: the task, type of data, and levels of the factors to name a few. The expansion of any of these areas is conceptually simple, but exponentially increases the number of participants needed to properly power the study. Additionally the sample of crowd-sourced, educated, but unfamiliar users may not extrapolate well to more experienced users. There are several ways that future work could be extended. Aside from expanding the support of the experimental factors, more exciting directions include: introducing a new task, including more visualizations, and changing the experience level of the target population. It is difficult to achieve good coverage given the number of possible factors to vary.

\section{Conclusion}\label{sec:conclusion}

This paper discussed a crowdsourced mixed design user study (\(n=108\)) comparing the efficacy of three linear projection techniques: PCA, grand tour, and radial tour. The participants performed a supervised cluster task, explicitly identifying which variables contribute to the separation of two target clusters. This was evaluated evenly over four experimental factors. In summary, mixed model regression finds strong evidence that using the radial tour sizably increases accuracy, especially when cluster separation location is not mixed at 33/66\%. The effect sizes on accuracy are large relative to the change from the other experimental factors and the random effect of data simulation, though smaller than the random effect of the participant. The radial tour was the most preferred of the three visuals.

There is no panacea for the comprehensive visualization of multivariate spaces. We have demonstrated that there is a definite value of user-control in linear projections. The agency of the analyst remains an important tool for the exploratory analysis of multivariate data.


\acknowledgments{
This research was supported by an Australian Government Research Training Program (RTP) scholarship. This article was created in \textbf{R} \cite{r_core_team_r:_2020} and \textbf{rmarkdown} \cite{xie_r_2018}. Visuals were prepared with \textbf{spinifex} \cite{spyrison_spinifex_2020}. We thank Jieyang Chong for his help in proofreading this article. The code, response files, their analyses, and the study application are publicly available at \url{https://github.com/nspyrison/spinifex_study}. The participant instruction video can be viewed at \url{https://vimeo.com/712674984}.}

\bibliographystyle{abbrv-doi}

\bibliography{spyrison-cook-marriott}
\end{document}